\documentclass[aps,floats,prl,showpacs,twocolumn]{revtex4-1}

\newcommand{\tp}{{\tilde \psi}}

\usepackage{amsfonts,amsmath} \usepackage{bm} \usepackage{dcolumn}
\usepackage{epsfig} \usepackage{latexsym}

\begin{document}

\title{Planck's quantum-driven integer quantum Hall effect in chaos}

\author{Yu Chen and Chushun Tian}

\affiliation{Institute for Advanced Study, Tsinghua University,
Beijing, 100084, P. R. China}



\begin{abstract}

The integer quantum Hall effect (IQHE) and chaos
are commonly conceived
as being unrelated.
Contrary to common wisdoms, we find in a canonical chaotic system, the kicked spin-$1/2$ rotor,
a Planck's quantum($h_e$)-driven phenomenon bearing a firm analogy to IQHE but of chaos origin.
Specifically, the rotor's energy growth is unbounded (`metallic' phase)
for a discrete set of critical $h_e$-values, but
otherwise bounded (`insulating' phase).
The latter phase is topological in nature and characterized by a quantum number (`quantized Hall conductance'). The number
jumps by unity whenever
$h_e$ decreases passing through each critical value.
Our findings, within the reach of cold-atom experiments, indicate that rich topological quantum phenomena may emerge from chaos.

\end{abstract}
\pacs{05.45.Mt,73.43.-f}
\maketitle

{\it Introduction.} While chaos is ubiquitous in nature, it is commonly conceived as being unrelated to the integer quantum Hall effect (IQHE) \cite{Klitzing80,Pruisken84a}, a unique phenomenon in two-dimensional ($2$D) electronic systems that heralded a revolution in condensed matter physics. Indeed,
the salient characteristic of chaos --
the extreme sensitivity
of system's behavior to small disturbances \cite{Gutzwiller90,Haake} --
is totally opposite to the robustness of the transport quantization protected by topological mechanisms \cite{Laughlin82,Thouless82}.
In addition, chaos occurs even in simple one-body systems.
For these systems, as we consider below, the concept of statistics
(Fermi, Bose, etc.) does not apply.
Yet, as recently reiterated by Raghu and Haldane \cite{Haldane08}, the Fermi statistics is an essential ingredient of IQHE:
integer fillings of Landau bands due to the Pauli principle give rise to
a quantized spectrum, $\mathbb{Z}$ (the integer set), of the Hall conductance.

The purpose of this Letter is to uncover a connection between these two seemingly incompatible subjects.
Serving as a prototype of many real complex systems, the kicked rotor is a canonical chaotic system \cite{Chirikov79a,QKR79,Izrailev90,Fishman84,Chirikov79,Fishman10}. We focus on it below.
Despite of its very simple construction -- a particle moving on a ring under the influence of sequential driving force (`kicking'), this system exhibits very rich phenomena, studies of which span a wide range of scientific disciplines (see Refs.~\cite{Chirikov79,Fishman10} for brief reviews).
Its realization in atom optics settings \cite{Raizen95} has triggered a renewal of studies of chaotic dynamics \cite{Zoller97,Ammann98,Hensinger01,Raizen01,Phillips06,Deland08,Chaudhury09}.

In chaotic systems, an important parameter governing the interplay between chaoticity and quantum interference is Planck's quantum, $h_e$, the ratio of
Planck's constant $\hbar$ to system's classical action. For kicked rotors, $h_e\equiv \tau \hbar/I$,
with $I$ the particle's moment of inertia and $\tau$ the kicking period, and a prominent feature of quantum chaoticity is the $h_e$-sensitivity of rotor's behavior. (Notably, for certain large $h_e$-values, namely, in deep quantum regime, kicked rotors exhibit even classical-like behavior \cite{Fishman03a,Wang11}.) This is a central yet highly challenging issue in studies of chaos \cite{Fishman10,Fishman84}.
A paradigm is that tuning $h_e$ may alter kicked rotor's symmetries, causing dramatic changes in system's behavior \cite{QKR79,Izrailev90,Izrailev80,Casati00,Altland10,Wimberger,Altland11,Altland14,Arcy01,Phillips06,Steinberg07}.


Here we report an exotic $h_e$-driven phenomenon totally beyond this paradigm: it is of {\it topological transition} nature \cite{footnote_1}. Specifically, we find that in quasiperiodically kicked spin-$1/2$ rotors there is an infinite number of topological quantum phases labeled by $\mathbb{Z}$, and lowering $h_e$ triggers sequential transitions between them.
Strikingly, although the system,
as simple as being one-body, one-dimensional ($1$D), and free of magnetic field,
is totally different from genuine quantum Hall systems (e.g., semiconductor heterojunctions subject to strong magnetic fields),
the sequential transitions between topological phases bear a firm analogy (in the sense of being mathematically equivalent) to IQHE. This is a novel-type IQHE, with $h_e^{-1}$ mimicking
the `filling fraction' (inversely proportional to applied magnetic field).
For this reason, it is dubbed as `the Planck's quantum-driven IQHE'. It suggests that rich topological quantum phenomena, without classical correspondence, may emerge from chaos.

{\it Main results.} The system is a spin-$1/2$ particle
moving on a ring subject to an external potential switched on at integer times \cite{footnote_2}.
The potential $V\equiv V_i(\theta_1,\theta_2+\tilde \omega t)\sigma^i$,
with $\theta_1$ the particle's angular position, $\theta_{2}$ an arbitrarily prescribed phase, and $\sigma^i,i=1,2,3$ Pauli matrices,
is modulated by
a frequency, $\tilde \omega$, incommensurate with the kicking frequency $2\pi$.
$V$ couples the rotor's spin and angular position degrees of freedom.
Such kind of kicking was first introduced by Scharf \cite{Scharf89}
and extensively studied later \cite{Bluemel94,Kravtsov04,Beenakker11}.
The main results are independent of the details of $V_i$ except that $V_i$ obey certain symmetry constraints (see below) and give rise to chaotic motion.

\begin{figure}[h]
\includegraphics[width=8.6cm]{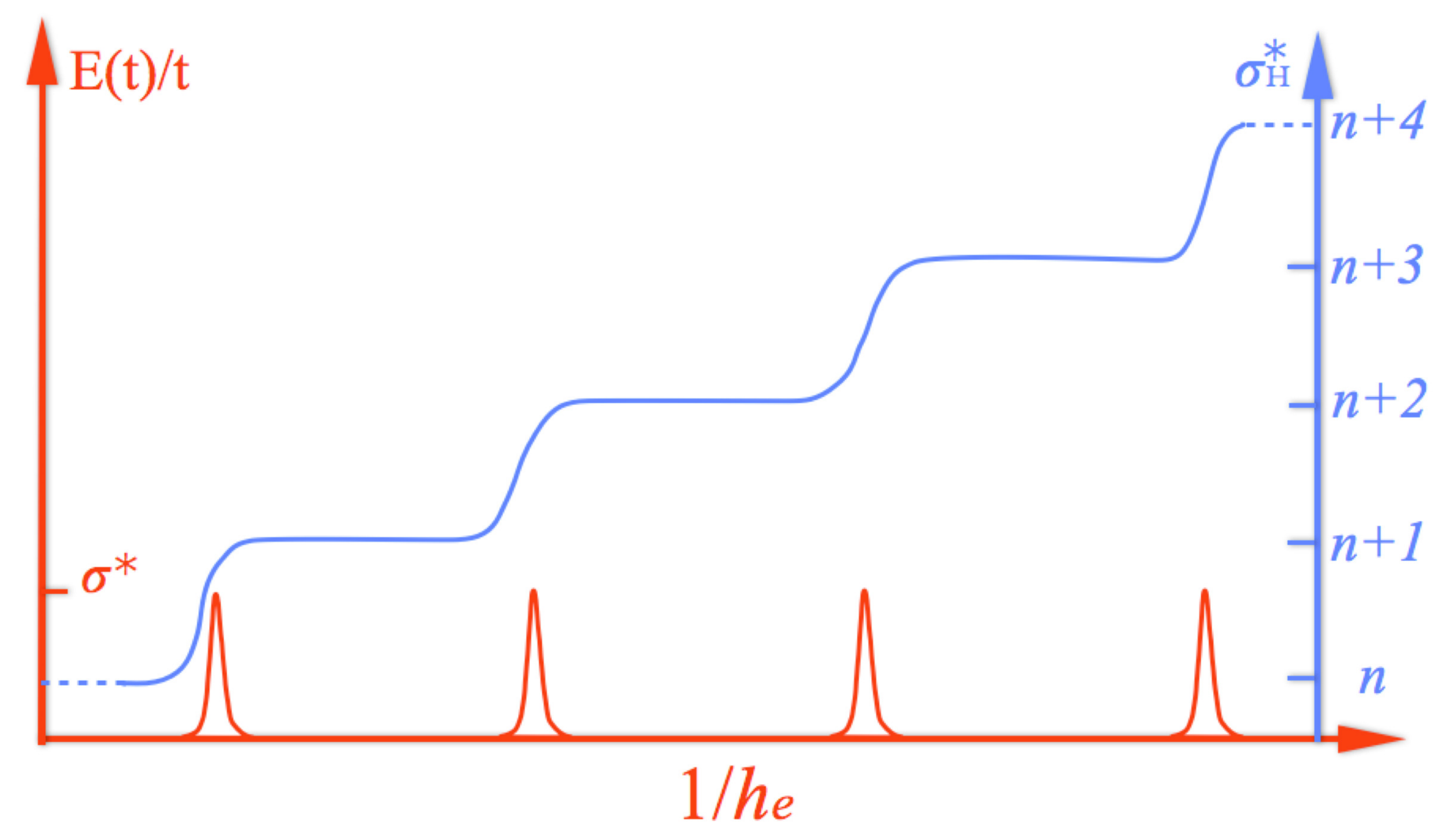}
\caption{Schematic representation of main results:
the red (blue) line is the profile of the
energy growth rate (the quantum number) versus
the inverse Planck's quantum.}
\label{fig:5}
\end{figure}

The particle's quantum motion is described by
\begin{eqnarray}
ih_e \partial_t \tp_t = {\hat H}(t) \tp_t\,,
\qquad\qquad\qquad \label{eq:1}\\
{\hat H}(t) =
-\frac{(h_e\partial_{\theta_1})^2}{2}+
V(\theta_1,\theta_2+\tilde \omega t)\sum_s\delta (t-s), \nonumber
\end{eqnarray}
with $\tp_t$ a two-component
spinor. As schematically shown in Fig.~\ref{fig:5}, this system exhibits rich quantum phase structures when $h_e^{-1}$ increases.
The phases are probed by the rotor's energy,
$ E(t) \equiv -\frac{1}{2}
\langle\!\langle \tilde \psi_t|\partial_{\theta_1}^2|\tilde \psi_t\rangle\!\rangle_{\theta_2}$,
with $\langle\cdot\rangle_{\theta_2}
$ the average over
$\theta_2$.
If the energy growth rate, $E(t)/t$,
is zero (non-zero) at long times, the rotor exhibits bounded (extended) motion in
angular momentum space and simulates an insulator (metal) in
disordered electronic systems. The red line shows that
there is an infinite discrete set of critical values of $h_e$.
At each critical value the rotor is a
metal with a universal
quantum energy growth rate,
$
\sigma^*
$, independent of system's details (e.g., $V_i$ and critical values),
while for other values rotor insulating states result.
Furthermore, we find that these insulating phases are topological, characterized by a
quantum number $\sigma_{\rm H}^*\in \mathbb{Z}$. The blue line shows that whenever $h_e^{-1}$ increases passing through each critical
value, $\sigma_{\rm H}^*$ jumps by unity.
Therefore,
the profiles of $E(t)/t
$ and
$\sigma_{\rm H}^*$ versus $h_e^{-1}$
bear a close analogy to those of IQHE: $h_e^{-1}$ mimics the filling fraction,
$E(t)/t
$ the longitudinal conductance, and $\sigma_{\rm H}^*$
the quantized Hall conductance. This analogy is put on a firm ground below,
and simulations of Eq.~(\ref{eq:1}) fully confirm our analytic predictions.

{\it Qualitative discussions.}
A possible physical interpretation of this phenomenon is as follows.
First of all, the
system can be traded for an equivalent (time-)periodic system. Indeed, by performing the transformation \cite{Casati89},
$\hat H \rightarrow e^{\tilde \omega t \partial_{\theta_2}}\hat H
e^{-\tilde \omega t\partial_{\theta_2}}$, $\tilde\psi_t
\rightarrow e^{-\tilde \omega t\partial_{\theta_2}} \tilde\psi_t \equiv \psi_t$ for Eq.~(\ref{eq:1}), we obtain $2$D autonomous stroboscopic dynamics,
\begin{equation}\label{eq:2}
    \psi_t = {\hat U}^t \psi_0, \quad
    \hat U \equiv e^{-\frac{i}{h_e}V(\theta_1,\theta_2)}e^{-\frac{i}{2}(h_e {\hat
n}_1^2+2{\tilde \omega} {\hat n}_2)},
\end{equation}
with $[\theta_{1,2},\hat n_{1,2}]=i$.
The first factor of $\hat U$ effecting a `spin-orbit coupling', with 
$(\theta_1,\theta_2)\equiv\Theta$ mimicking the usual momentum; the second oscillates rapidly in $n_{1,2}$ for generic values of $h_e$ (and $\tilde \omega$ incommensurate with $2\pi$), generating an effective `strongly disordered potential'.

Then, the `disordered potential' causes 
localization in $2$D angular momentum space [with coordinate $(n_1,n_2) \equiv N$].
Since a Chern character $\in \mathbb{Z}$ can in principle be constructed even for such strongly inhomogeneous systems (with broken time-reversal symmetry),
following the
mathematical framework of Bellisard {\it et. al.} \cite{Bellisard94}, the rotor insulator is topological. On the other hand, the oscillatory factor is not genuine disorders. In general, tuning $h_e$ gives rise to a variety of disorder structures -- a profound consequence of quantum chaoticity. Provided the potential generated includes an extended equipotential line, a velocity transverse to the potential gradient, arising from the spin-orbit coupling \cite{Niu10}, then drives an unbounded motion along this line. This is a hallmark of delocalization. Because the Chern character is well defined only if the motion is bounded \cite{Bellisard94}, a plateau transition occurs simultaneously.

{\it Microscopic theory.} Now we outline the proof. The details will be published elsewhere \cite{Tian14}. We consider $V_i(\theta_1,\theta_2)$ that are
$2\pi$-periodic in $\theta_{1,2}$ and invariant upon exchange of the two variables.
Moreover, $V_{1}$ has odd (even) parity in
$\theta_1 (\theta_2)$ and {\it vice versa} for $V_{2}$;
$V_3$ has even parity in both. Note that $V_3$
breaks the time-reversal symmetry $
i\sigma^2 K$,
where $K$ is the combination of complex conjugation
and the operation: $t\rightarrow -t, \theta_{1,2}\rightarrow -\theta_{1,2}$.

\begin{figure}[h]
\includegraphics[width=8.6cm]{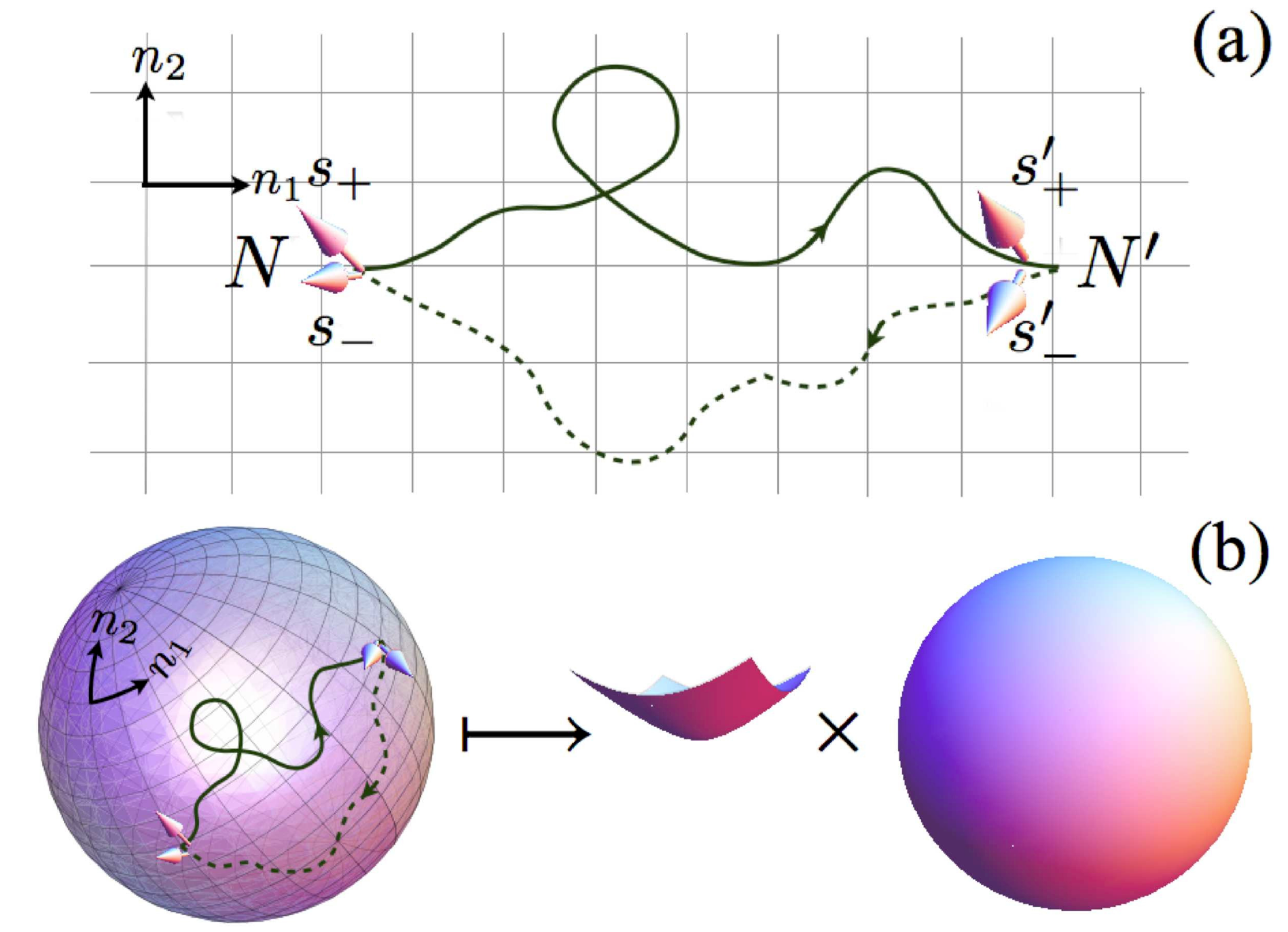}
\caption{
(a) The
$2$D autonomous stroboscopic dynamics (\ref{eq:2}) is governed by interference between
the advanced (solid line) and the retarded (dashed line) quantum amplitudes.
(b) Mappings
from the $2$D angular momentum space,
with its boundary identified as a point,
onto the supersymmetry $\sigma$-model space of unitary symmetry $\simeq$
$2$-hyperboloid $\times\,2$-sphere.
}
\label{fig:1}
\end{figure}

We introduce the function,
$K_\omega(Ns_+s_-,N's'_+s'_-)$, defined as $\langle\!\langle Ns_+|\frac{1}{1-e^{i\omega_+}{\hat U}}|N's_+'\rangle
\langle N's'_-|\frac{1}{1-e^{-i\omega_-}{\hat U}^\dagger}|Ns_-\rangle\!\rangle_{\omega_0}$
to characterize the motion (\ref{eq:2}),
where $s_\pm,s'_\pm$ are spin indices, $\omega_\pm\equiv \omega_0\pm\frac{\omega}{2}$ with
$\omega$ understood as $\omega+i0$, and $\langle\cdot\rangle_{\omega_0}$ is the average
over the quasi-energy spectrum.
Physically, $K_\omega$ describes interference between
the advanced and the retarded quantum amplitudes (Fig.~\ref{fig:1}a).
It gives the energy as
$E(t)=\frac{1}{2}{\rm Tr} (\hat n_1^2 K_\omega {\psi}_0\otimes {\psi}_0^\dagger)$,
where the trace `Tr' includes the angular momentum and spin indices
of the theory,
and $\psi_0$ is uniform in
$\Theta
$-representation.

Similar to usual kicked (spinless) rotors \cite{Altland10,Altland11},
$K_\omega$ can be written as a functional integral over
$Q\equiv \{Q_{NN'}\}$, where $Q_{NN'}$ is a $4\times 4$ supermatrix.
The off-diagonality of $Q$ in angular momentum space ($N\neq N'$)
encodes information about
the `momentum' $\Theta$ of the coherent propagation of
the advanced and the retarded quantum amplitudes.
Due to chaoticity, the memory on $\Theta$ is
lost. Consequently, the off-diagonality is eliminated, yielding $Q_{NN'}\equiv
Q(N)\delta_{NN'}$.
$Q(N)$
entails mappings from the $2$D angular momentum space (with its boundary identified as a point) onto
the $\sigma$-model space \cite{Efetov97},
$\frac{U(1,1)}{U(1)\times U(1)} \times \frac{U(2)}{U(1)\times U(1)}$,
where the first (second) factor is $\simeq$ $2$-hyperboloid(sphere)
(Fig.~\ref{fig:1}b). The corresponding homotopy group is $\mathbb{Z}$.
This justifies the topological nature of the Planck's quantum-driven IQHE.
We emphasize that this nontrivial homotopy group
arises only if chaotic motion develops in all directions of $2$D angular momentum space; otherwise the mapping above would not exist.
(In fact, we have found that even partial restoration of regular dynamics
leads to the disappearance of the novel-type IQHE.)

The details of
the $2$D motion (\ref{eq:2}) at large time scales ($\omega \ll 1$) are encoded by the effective action,
\begin{eqnarray}\label{eq:111}
    S[Q]\!=\!{1\over 4} \mathrm{Str}\!\left(\!-\sigma
    (\nabla Q)^2+\sigma_{\rm H}
Q \nabla_1 Q \nabla_2 Q
  -2i\omega Q\Lambda\!\right),\quad
\end{eqnarray}
where $Q(N)$ smoothly varies in $N$ and the supertrace `Str' includes the angular momentum.
The bare (unrenormalized) coefficients are given by
\begin{eqnarray}
\sigma &=& 2\int\!\!\!\!\int \frac{d\theta_1}{2\pi}\frac{d\theta_2}{2\pi}
    \frac{\partial_{\theta_1}\!\epsilon_i\partial_{\theta_1}\!\epsilon_i}{(\epsilon^2+1)^2},
    \label{eq:3}\\
\sigma_{\rm H} &=& 4\varepsilon^{ijk} \int\!\!\!\!\int \frac{d\theta_1}{2\pi}\frac{d\theta_2}{2\pi}
\hat O\left(\frac{\partial_{\theta_1}\!\epsilon_j
\partial_{\theta_2}\!\epsilon_k}{(\epsilon^2+1)^2}\right),
\label{eq:67}
\end{eqnarray}
where $\epsilon_i=\cot\frac{|V|}{2h_e}\frac{V_i}{|V|}$, $\varepsilon^{ijk}$ is the totally antisymmetric tensor,
and the operator $\hat O\equiv \epsilon_i+ \int d\mu \partial_\mu\epsilon_i$.
In deriving these results we note that $\hat U$ acts on a Hilbert space composed of
wave functions which vanish at infinity. To satisfy this condition, we allow
$V$ to depend smoothly on certain parameter, $\mu$,
on a boundary strip of angular momentum space. This leads to the $\mu$-dependence of $\epsilon_i$
and the $\mu$-integral in $\hat O$.

Equation (\ref{eq:111})
gives a Pruisken-type field theory
for IQHE \cite{Pruisken84a,Pruisken84b},
with Eqs.~(\ref{eq:3}) and (\ref{eq:67})
mimicking the unrenormalized longitudinal and Hall conductances,
respectively. It justifies that the rotor metal-insulator 
transition at critical $h_e$-values is in the same universality
class as the integer quantum Hall transition.
The second term in $S[Q]$
is a topological $\theta$-term. Without this term the effective field theory is identical to
that describing $2$D Anderson localization of quasiperiodically kicked rotors \cite{Altland11},
and a rotor insulator results irrespective of $h_e$.

\begin{figure}[h]
\includegraphics[width=8.6cm]{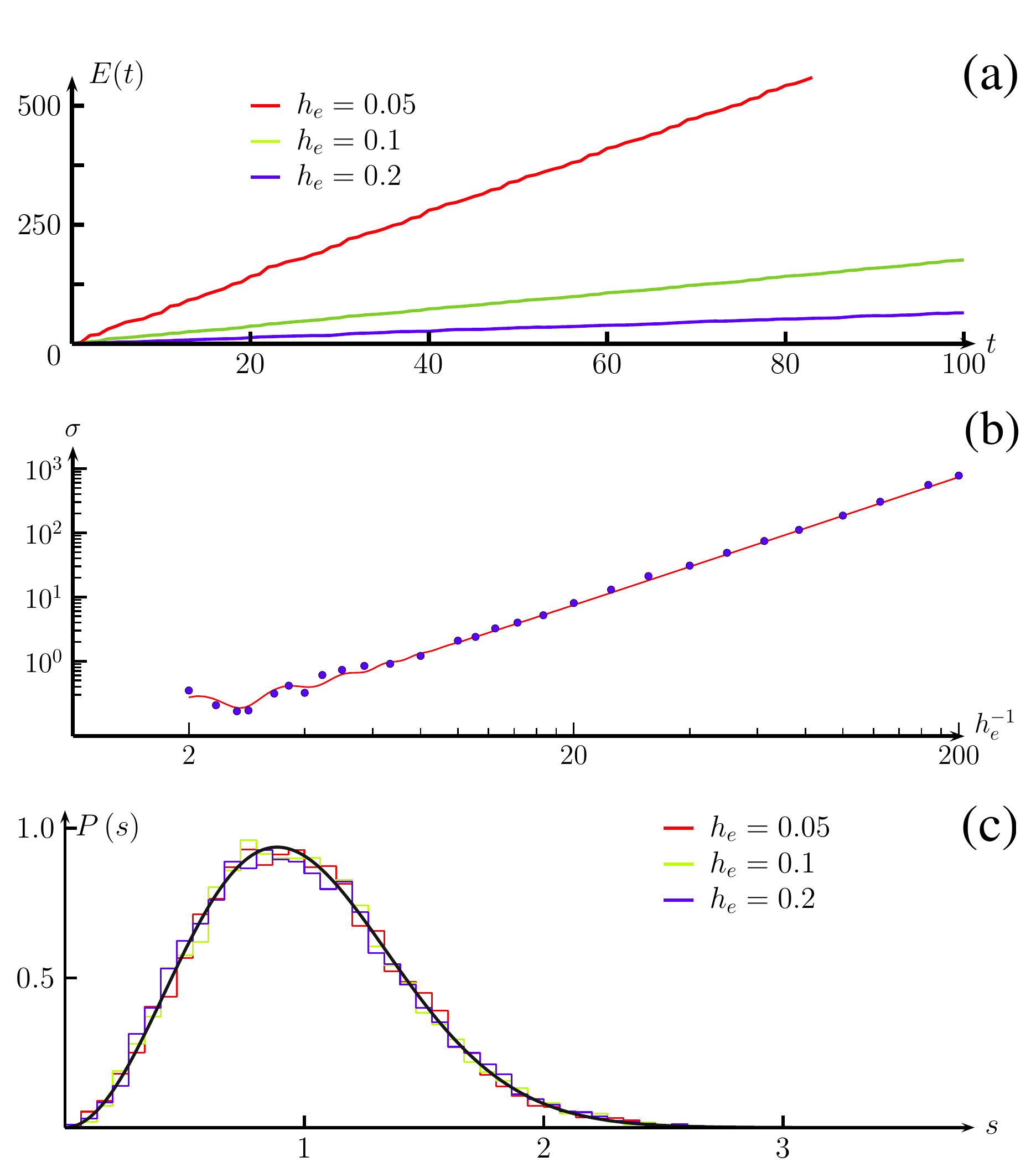}
\caption{
(a) Simulations of Eq.~(\ref{eq:1}) show that at short times
the rotor's energy grows linearly. (b) The numerical (circles) and the analytical (solid line)
results for the energy growth
rate are in excellent agreement. (c) The Floquet operator governing the $2$D autonomous stroboscopic dynamics (\ref{eq:2}) has a chaotic quasi-energy spectrum,
with the level spacing distribution (histograms) obeying the Wigner surmise
(solid line).}
\label{fig:2}
\end{figure}

By expressing the function, $\sum_{ss'}K_\omega(Nss,N's's')$, as a functional integral
over the $Q$-field,
we find that the Fourier transformation (with respect to $N-N'$) of this function
is $\sim (-i\omega + \sigma(\omega) \varphi^2)^{-1}$ and exhibits a diffusive pole,
where $\varphi$ is the Fourier wavenumber, and $\sigma(\omega)$ simulates the `optical conductance' in condensed matter.
This pole leads to an important general relation:
$-\sigma(\omega)/\omega^2$ is the Fourier transformation of $E(t)$.
Using perturbative methods, one finds
$\sigma(\omega)=\sigma$. As a result, $E(t)=\sigma t$. Physically, this implies that in early times
($t\ll e^{\sigma^2}$) interference
plays no roles and chaotic diffusion in angular momentum space dominates (cf. Fig.~\ref{fig:2}a).

\begin{figure}[h]
\includegraphics[width=8.6cm]{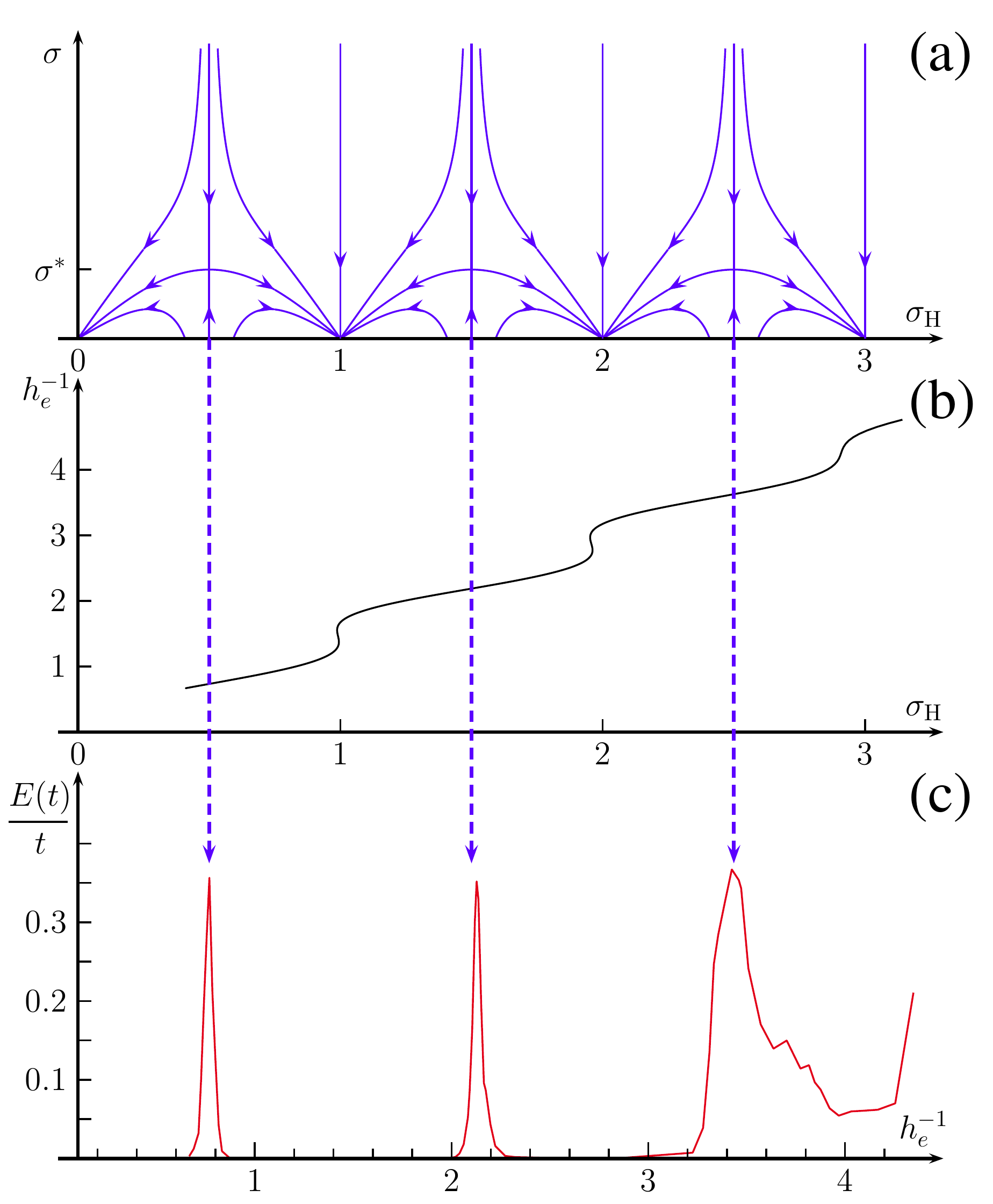}
\caption{
In conjunction with the renormalization group flow (a),
Eq.~(\ref{eq:67}) (solid line) predicts
three transitions (the intersections of solid and dashed lines)
when $h_e$ varies from $0.23$ to $1.50$, i.e., $0.67 \leq h_e^{-1}\leq 4.35$
(b). Long-time simulations confirm this,
with the critical values in agreement with analytic predictions;
they further confirm that the rotor is metallic
at the transition with a universal quantum growth rate
and otherwise is insulating with a vanishing growth rate
(c).}
\label{fig:3}
\end{figure}

To study $E(t)$ at long times (or for small $\sigma$) we resort to non-perturbative renormalization group analysis \cite{Pruisken84a,Pruisken84},
obtaining a two-parameter renormalization group flow
(Fig.~\ref{fig:3}a).
For this flow there are two types of fixed points:
$(\sigma_{\rm H}^*=n\in \mathbb{Z},0)$ and $(n+\frac{1}{2},\sigma^*)$.
The former is stable, with a vanishing zero-frequency conductance
which is a main characteristic of insulators
and implies that $E(t)$ saturates at long times.
These insulators are conceptually different from usual rotor insulators
\cite{QKR79,Izrailev90,Chirikov79,Fishman10,Casati89,Raizen95,Altland10,Altland11,Deland08,Fishman84}:
they are characterized by a quantum number,
$\sigma_{\rm H}^*$, that is endowed with topological nature by the $\theta$-term of $S[Q]$.
The latter type of fixed points is stable along the critical line
of $\sigma_{\rm H}=n+\frac{1}{2}$ but unstable in the $\sigma_{\rm H}$-direction,
with $\sigma(\omega\rightarrow 0)=\sigma^*$
which is a main characteristic of metals and gives $E(t)=\sigma^*t$ for long times.
This metal has a small but universal energy growth rate $\sigma^*={\cal O}(1)$
and is of quantum nature.

From Eq.~(\ref{eq:67}) we find for $h_e\ll 1$
\begin{equation}\label{eq:6}
    \sigma_{\rm H} \propto h_e^{-1},
\end{equation}
with the proportionality coefficient
depending merely on $V_i$. Since $\sigma_{\rm H}$ increases unboundedly with $h_e^{-1}$,
there is an infinite discrete set of critical $h_e^{-1}$-values for which
$\sigma_{\rm H}$ is a half-integer and the system is at the critical line.
When $h_e^{-1}$ falls between two nearest critical values,
the renormalization group flow
drives the system to the same insulating phase,
giving rise to a plateau in Fig.~\ref{fig:5}.
Whenever $h_e^{-1}$ increases passing through each critical value,
$\sigma_{\rm H}^*$ jumps by unity (the plateau transition) and simultaneously,
the system exhibits a metal-insulator transition (see Fig.~\ref{fig:5}). Moreover,
because of Eq.~(\ref{eq:6}) large critical $h_e^{-1}$-values are equally spaced along the $h_e^{-1}$-axis.

{\it Numerical confirmation.} For simulations we use $V=\frac{2\arctan 2d}{d} \boldsymbol{d}\cdot\boldsymbol\sigma,
\boldsymbol{d}=(\sin\theta_1, \sin\theta_2, 0.8(\mu-\cos\theta_1-\cos\theta_2))$.
It was previously \cite{Beenakker11} used to find
a $\mu$-driven ($h_e$ set to unity) effect analogous to that predicted by Qi {\it et. al.}\cite{Qi06}. Here, because we are interested in the $h_e$-dependent physics,
we fix $V$ and set $\mu=1$.

We use standard fast Fourier transform techniques to simulate the $1$D evolution (\ref{eq:1}) with $\tilde \omega=\frac{2\pi}{\sqrt{5}}$.
The simulation results show that at short times $E(t)$
grows linearly (Fig.~\ref{fig:2}a). The growth rate,
$E(t)/t$, is in excellent agreement with the analytical formula
(\ref{eq:3}) for $5\times 10^{-3} \leq h_e \leq 5\times 10^{-1}$ (Fig.~\ref{fig:2}b).
The results are averaged over $10^2$ values of $\theta_2$.
This linear growth
is a manifestation of chaotic motion in angular momentum space.
To further confirm the chaoticity, we study the equivalent $2$D system (\ref{eq:2}).
Specifically, we calculate the quasi-energy spectrum of $\hat U$
by numerical diagonalization.
The Hilbert space is composed of wave functions
on the angular momentum lattice of $64\times 64$ sites
satisfying periodic boundary conditions.
We find that the quasi-energy spectrum is chaotic
with the level spacing distribution, $P(s)$, obeying the Wigner surmise for
the circular unitary ensemble \cite{Haake} (Fig.~\ref{fig:2}c).

Next, we simulate the long-time ($t\leq 6\times 10^5$) evolution (\ref{eq:1}) for $0.23\leq h_e \leq 1.50$.
For this range of $h_e$
Eq.~(\ref{eq:67}) (the solid line in Fig.~\ref{fig:3}b), in conjunction with the renormalization group flow, 
predicts four insulating phases, corresponding to
the fixed points of $\sigma_{\rm H}^*=0,1,2,3$, respectively,
and three transitions at $h_e^{-1}=0.73,2.19$, and $3.60$
(the intersections of the solid and the three dashed lines
in Fig.~\ref{fig:3}b),
corresponding to the critical lines of $\sigma_{\rm H}=\frac{1}{2},\frac{3}{2}$, and $\frac{5}{2}$, respectively.
Figure \ref{fig:3}c is the simulation results of $E(t)/t$ at $t=6\times 10^5$.
It exhibits three sharp peaks (transitions)
at $h_e^{-1}=0.77, 2.13$, and $3.45$, in excellent agreement with
analytic predictions. The peaks are approximately uniformly spaced with a spacing
$\approx 1.34$, in agreement with the prediction of Eq.~(\ref{eq:6}).
The peak values are approximately the same, $E(t)/t\approx 0.35$,
signaling a rotor metal with universal quantum growth rate. Between the peaks,
$E(t)/t$ is fully suppressed indicating a rotor insulator.
Thus, simulations fully confirm the analytic predictions for the novel-type IQHE.

{\it Conclusion.} We have shown analytically and confirmed numerically
a chaos-induced IQHE in
a simple -- $1$D, one-body, and free of magnetic field -- system.
This novel-type IQHE exists in a large class of driven chaotic systems. 
It suggests that rich topological quantum phenomena may emerge from chaos.
We expect its confirmation to be within the reach of cold-atom experiments
\cite{Deland08}, where realizing the system (\ref{eq:1}) and tuning-$h_e$ seem to be possible.

Many intriguing questions are opened.
Notably, would there be any counterparts of the fractional quantum Hall effect in chaotic systems?
In addition, are there any relations between the present rotor insulator and the Floquet topological insulator found in driven but {\it non-chaotic} systems \cite{Galitski11}?
Finally, the rotor insulator, lack of translation symmetry, resembles the topological Anderson insulator
in disordered systems \cite{Altland14a}.
Our findings may bring a new angle to this currently active subject.

We are grateful to I. Guarneri, G. Casati, and J. Wang for stimulating discussions and to A. Kamenev for useful comments on the manuscript.
Work supported by the NSFC (No. 11174174) and by the Tsinghua Univ. ISRP.


\begin{thebibliography}{}
\bibitem{footnote_1} Throughout this work we do not consider the nongeneric case
of rational $h_e/(4\pi)$.
\bibitem{footnote_2} We rescale the time by $\tau$ and the rotor's angular momentum by $I/\tau$.
\end{thebibliography}
\end{document}